\documentclass[twocolumn,superscriptaddress]{revtex4}

\usepackage{graphicx}

\begin{document}

\title{Phonon-bottleneck enhanced magnetic hysteresis in a molecular paddle wheel complex of Ru$_2^{5+}$}

\author{L. Chen}
\affiliation{Department of Physics, Florida State University, Tallahassee, Florida 32310 and National High Magnetic Field Laboratory, Florida State University, Tallahassee, Florida 32310}

\author{C.M. Ramsey}
\affiliation{Department of Chemistry and Biochemistry, Florida State University, Tallahassee, Florida 32306 and National High Magnetic Field Laboratory, Florida State University, Tallahassee, Florida 32310}

\author{N.S. Dalal}
\affiliation{Department of Chemistry and Biochemistry, Florida State University, Tallahassee, Florida 32306 and National High Magnetic Field Laboratory, Florida State University, Tallahassee, Florida 32310}

\author{T. Ren}
\affiliation{Department of Chemistry, Purdue University, West Lafayette, Indiana 47907}

\author{F.A. Cotton}
\affiliation{Department of Chemistry, Texas A\&M University, College Station, Texas 77842}

\author{W. Wernsdorfer}
\affiliation{Laboratoire Louis N\'{e}el-CNRS, BP 166, 38042 Grenoble, France}

\author{I. Chiorescu}
\altaffiliation{Electronic email: ichiorescu@fsu.edu}
\affiliation{Department of Physics, Florida State University, Tallahassee, Florida 32310 and National High Magnetic Field Laboratory, Florida State University, Tallahassee, Florida 32310}

\begin{abstract}
The ruthenium based molecular magnet [Ru$_2$(D(3,5-Cl$_2$Ph)F)$_4$Cl(0.5H$_2$O)$\cdotp$C$_6$H$_{14}$] (hereafter Ru$_2$) behaves as a two-level system at sufficiently low temperatures. The authors performed spin detection by means of single-crystal measurements and obtained magnetic hysteresis loops around zero bias as a function of field sweeping rate. Compared to other molecular systems, Ru$_2$ presents an enhanced irreversibility as shown by ``valleys'' of negative differential susceptibility in the hysteresis curves. Simulations based on phonon bottleneck model are in good qualitative agreement and suggest an abrupt spin reversal combined with insufficient thermal coupling between sample and cryostat phonon bath.

\end{abstract}

\date{Received 11 July 2006; accepted 14 November 2006: APL, \textbf{89}, 252502 (2006)}

\maketitle

Molecular magnets have been explored intensively in the recent years for their potential application in information technology and as test beds for quantum mechanics at macroscopic scale. These systems are tunable, identical, two- or multilevel systems that can be produced in large numbers and therefore are potential candidates for qubit implementation in quantum computing algorithms \cite{qcmn12}. In crystalline form, the molecules are relatively well isolated from each other and thus the quantum physics of their spin is not strongly affected by the collective nature of the sample. Large spin molecular magnets have demonstrated key quantum effects, such as quantum tunneling of the magnetization through an anisotropy barrier \cite{qtm,qtm2} and Berry phase interference \cite{berryFe8}. Low spin molecules have shown interesting effects when tuning phonon dissipation \cite{v15b} that provide increased relaxation and decoherence times for the molecular qubits.

\begin{figure}
\includegraphics[width=8cm]{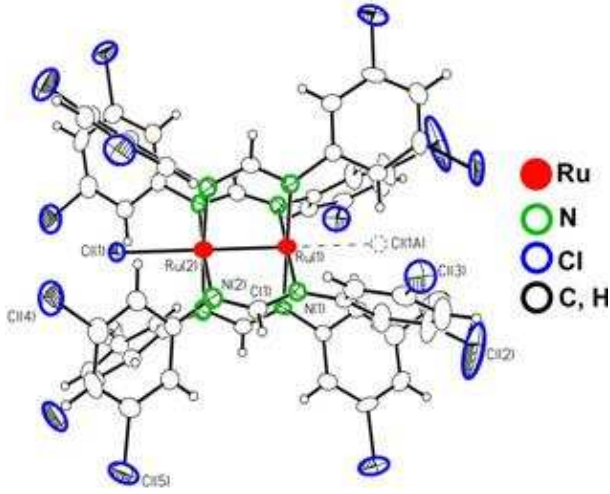}
\caption{The [Ru$_2$(D(3,5-Cl$_2$Ph)F)$_4$Cl(0.5H$_2$O)$\cdotp$C$_6$H$_{14}$] molecule with nondisordered atoms shown as displacement ellipsoids at the $30\%$ probability level (H atoms are omitted for clarity). The two Ru ions form the magnetic core of the molecules, synthesized in a single crystal with tetragonal symmetry. The Ru-Ru axes of all molecules are parallel to one another and aligned with the $c$ axis (001) of the unit cell.}
\label{fig1}
\end{figure}

In the case of low-spin molecular systems, the anisotropy barrier against spin reversal is often very small. However, magnetic hysteresis has been recently measured for a series of low-spin molecules (V$_{15}$ S=1/2, Fe$_{10}$ S=1, V$_{6}$ S=1/2) \cite{v15a,v15b,Fe10,V6} and explained in the frame of the phonon-bottleneck (PB) model \cite{AB,v15a}. In this model, the phonons' low heat capacity keeps the lattice and the spins at the same temperature $T_s$, slightly different from cryostat temperature $T$, as a result of inherent limited sample thermalization in the experimental setup. Cycling of an external magnetic field induces a delay (hysteresis) in the dynamics of the spin temperature which is observed by measuring the magnetic moment of a single crystal. Other theoretical interpretations treat the lattice at equilibrium with the cryostat at a temperature $T\neq T_s$ \cite{equil}. Though magnetic hysteresis can be simulated for a given bath temperature, these models cannot explain the observed dependence on sample thermalization \cite{v15b} and the picture of phonons having very low heat capacity at low temperatures \cite{AB}. On the other hand, the PB model presented in Ref.~\cite{v15a} gives a good quantitative agreement with the experiments. The present study introduces a particular aspect of the phonon-bottleneck phenomena as observed in the Ru$_2$ paddle wheel molecular system (structure in Fig.~\ref{fig1}).  Single crystal measurements reveal enhanced magnetic irreversibility with ``valleys'' of negative differential susceptibility followed by rapid increase of magnetization. The observed dynamics can be explained quantitatively in the frame of the PB model for an abrupt spin reversal in zero field, followed by a rapid gain in phonon energy.

\begin{figure}
\includegraphics[width=8cm]{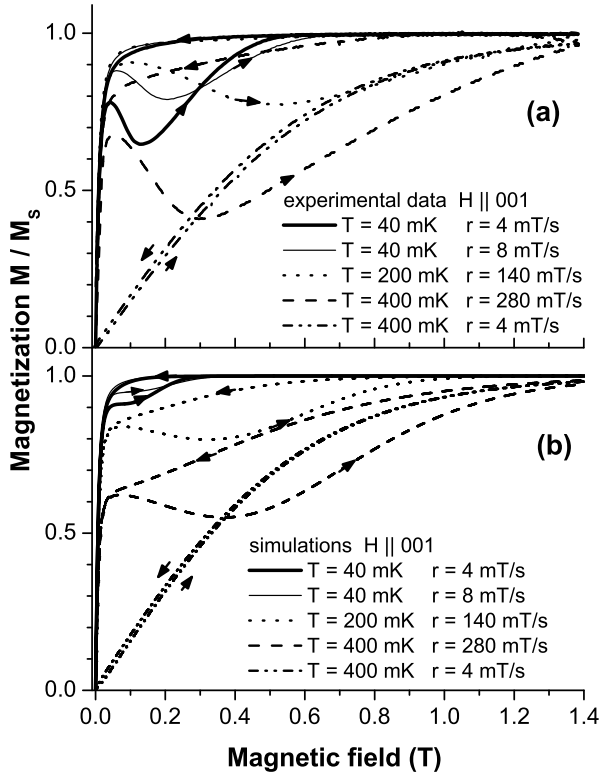}
\caption{Experimental (a) and simulated (b) hysteresis curves for a magnetic field along the 001 axis of the sample measured at various temperatures and magnetic field sweeping rates.}
\label{fig2}
\end{figure}

The [Ru$_2$(D(3,5-Cl$_2$Ph)F)$_4$Cl(0.5H$_2$O)$\cdotp$C$_6$H$_{14}$] molecular crystal was synthesized following the methods described in Ref.~\cite{Ru2}. It contains a magnetic Ru$_2^{5+}$ core with Ru$-$Ru bond length of 2.36~$\textrm{\AA}$, surrounded by bridging ligands in a paddle wheel pattern. Each molecule is formed of a dinuclear core which allows the ground state of the molecule to be described as a single spin. The crystals are tetragonal, and the Ru-Ru axes of all molecules are parallel to one another and aligned with the $c$ axis (001) of the unit cell. The paramagnetic units are separated by 11.2~$\textrm{\AA}$ (nearest neighbor distance) so that the dipolar interaction between molecules is very weak. The compound can be synthesized in relatively large crystal sizes with low-index faces suitable for single crystal experiments. Magnetic resonance studies \cite{Ru2} indicate a total spin $S=3/2$ with the lowest two levels $S_z=\pm1/2$ located $\sim200$~K below the $S_z=\pm3/2$ levels. Consequently, the Zeeman splitting in a longitudinal field corresponds to a spin 1/2 system, whereas in a transverse field it is twice as large due to increased mixing with the upper levels\cite{Ru2}.

The effective Hamiltonian of a two-level system can be written as $H_{eff}=-\epsilon\sigma_z/2-\Delta_0\sigma_x/2$, where $\sigma_{z,x}$ are the Pauli spin matrices, $\Delta_0$ is the level repulsion, and $\epsilon=g\mu_BH$ is the spin bias ($\mu_B$ is the Bohr magneton and $H$ is the external magnetic field, always $\parallel$ to z-axis of $H_{eff}$). The above mentioned effect of a transverse field on the Zeeman energy can be described by a gyromagnetic factor $g\cong2$ when $H\parallel001$ and $g\cong4$ when $H\parallel110$. Thus, the energy separation can be written as $\Delta_H=\sqrt{\Delta_0^2+\epsilon^2}$, which is also the energy $\hbar\omega=\Delta_H$ of phonons responsible for spin thermalization.

\begin{figure}
\includegraphics[width=8cm]{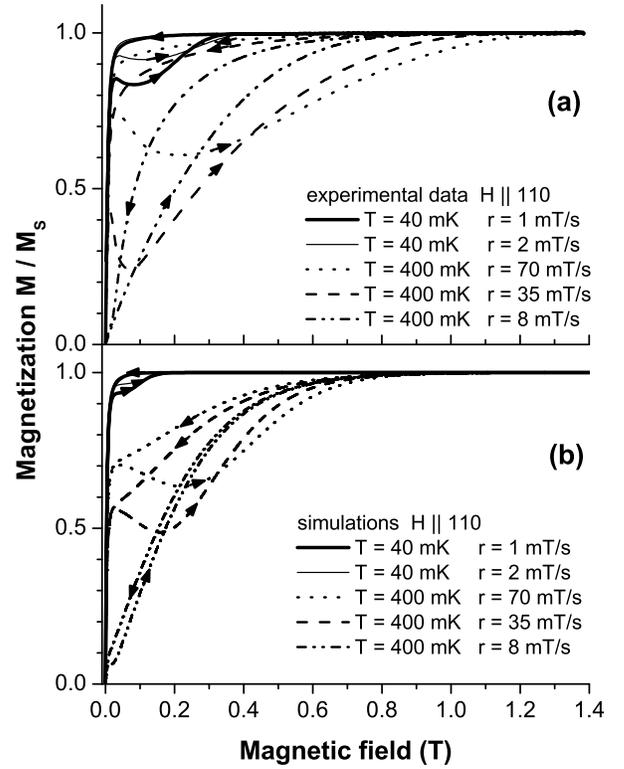}
\caption{Hysteresis curves for a magnetic field along the 110 axis of the sample: (a) experimental data measured at various temperatures and magnetic field sweeping rates;  (b) simulated cycles in the same conditions as in (a).}
\label{fig3}
\end{figure}

To analyze the role of phonon bottleneck on the spin dynamics of the Ru$_2$ molecule, we performed single-crystal magnetic measurements using a micron-sized superconducting quantum interference device \cite{qtm2} at very low temperatures ($\lesssim$1~K) when the molecule is purely a two-level system. Hysteresis cycles in field swept between $\pm$1.4~T at constant rate are presented in Figs.~\ref{fig2}a,\ref{fig3}a (for clarity, only one half of the cycle is shown; the other half being symmetric against the origin). As in the case of other low-spin molecules \cite{v15a,v15b,Fe10,V6}, an irreversibility is occurring even though no anisotropy barrier against spin reversal was detected.

In the case of the Ru$_2$ molecule we notice a large spin flip during passage through the origin, as shown by an abrupt ascending branch that deviates from the reversible behavior and can reach $\sim$90$\%$ of the total magnetization (see data for $T=40$~mK, $r=8$ and 2~mT/s in Figs.~\ref{fig2}a and \ref{fig3}a respectively). This is the signature of a small level repulsion $\Delta_0$, which in the absence of thermal broadening and for sufficiently small spin temperatures is observed as an abrupt spin flip (see also Ref.~\cite{V6}). We note that the nearly full spin reversal is still done in a quantum adiabatic fashion because of fairly low sweeping rate (no Landau-Zener excitations). 

When in thermal equilibrium and for $\Delta_0$ large enough to prevent nonadiabatic excitations, the molecule spins follow $H$ adiabatically through zero field showing a \emph{spin up} - \emph{spin down} transition, as observed at low sweep rates (see Fig.~\ref{fig2} for T=0.4~K and $r=4$mT/s). However, for sufficiently large sweeping rates the number of available phonons at energy $\Delta_H(H)$ can decrease if sample thermalization is not perfect. In Fig.~\ref{fig3} a small effect is visible already at T=0.4~K and $r=8$mT/s. At low temperatures phonons have much less heat capacity than spins \cite{AB} and the spin-phonon coupled system can be treated as being at the same temperature $T_s$, defined by $n_1/n_2=\exp(\Delta_H/k_BT_s)$, with $n_{1,2}$ the out of the equilibrium occupation numbers of the ground and excited spin levels, respectively. Our experiments show a steep decreasing field branch [see arrows for the first four curves in Figs.~\ref{fig2}a and \ref{fig3}a] indicating $T_s<T$, where $T$ is the cryostat temperature. This behavior is due to a lack of thermal coupling between the lattice and the phonon reservoir of the cryostat leading to an insufficient number of available phonons for spin thermalization. This behavior persists after passing through zero field as well. Quantitatively, this phenomenon can be described by the PB model \cite{AB,v15a} in which deviations from equilibrium are given by $\dot{x}\tau_H=1/x-1$ with $x=(n_1-n_2)/(n_{1eq}-n_{2eq})$ and spin relaxation time \cite{AB}:
\begin{equation}
\tau_H=\alpha\frac{\tanh^2(\Delta_H/2k_BT)}{\Delta_H^2}.
\end{equation}
The parameter $\alpha=2\pi^2\hbar^2v^3N\tau_{ph}/3\Delta\omega$ is proportional to molecule density $N$, phonon velocity $v$ and phonon-bath relaxation time $\tau_{ph}$ (depending on the sample thermalization to the external bath) and inversely proportional to the spectral width of the levels $\Delta\omega$.

In contrast to other molecular systems, the Ru$_2$ spin-flip induces a large increase of phonon energy that is visible through a valley of negative differential susceptibility. The magnitude of the effect and its dynamics depend on the competition between two field-dependent parameters, the relaxation rate $\tau_H$ and the time $\Delta_H/v$ spent at a given two-level energy separation (with $v=d\Delta_H/dH$). The PB model can describe accurately this phenomenon as one can see from the simulations of Figs.~\ref{fig2}b and \ref{fig3}b. Following the abrupt spin reversal, the spin system is ``scanning" again the phonon energy distribution and available resonant phonons are used to excite the spins. Subsequent spin deexcitations will generate phonons with slightly higher energy since $\Delta_H$ has been increased meanwhile. In a short time, these dynamics increase the phonon energy density which compromises the magnetization rapidly and stimulates emission of further phonons \cite{AB}. The gain in phonon energy, combined with the imperfect thermal coupling of the sample, leads to an increased spin temperature. Thus, in this field range the phonon bottleneck acts in the opposite way, that is, the phonons cannot ``escape" from the sample into the external bath.

The simulations using the PB model require two fit parameters, namely, the zero-field level repulsion $\Delta_0$ and the sample thermalization parameter $\alpha$. We find $\Delta_0=15$~mK and $\alpha=1$ and 3~sK$^2$ for $H\parallel001$ and $H\parallel110$, respectively. The measurements of Figs.~\ref{fig2}a and \ref{fig3}a were done in two separate runs with inherently different sample thermalizations to the cryostat. The resulting different values of $\alpha$ indicate an intermediate thermalization. As a comparison, in related experiments on V$_{15}$ molecule \cite{v15b}, $\alpha$ was 0.09 and 130~sK$^2$ for a sample well thermalized and well isolated, respectively, from the cryostat. 

The obtained value of $\Delta_0$ is several times smaller than the value found for $V_{15}$ ($\Delta_0=80$~mK \cite{v15b}). This causes a more complicated situation near zero field, where the spin temperature is strongly reduced. In this case, the PB model assumption $T_s\lesssim T$ is no longer valid, which affects the agreement between simulations and experiments.

In conclusion, we introduced a particular aspect of the phonon-bottleneck phenomena as detected by single-crystal magnetic measurements of the Ru$_2$ paddle wheel molecular system. Ru$_2$ is a member of a class of molecular magnets with a very simple (diatomic) core that exhibits this phenomenon; all the previous cases involved much more complex spin metal clusters. Enhanced magnetic irreversibility with ``valleys'' of negative differential susceptibility followed by rapid increase of magnetization is observed. These dynamics are due to an abrupt spin reversal in zero field, followed by an increase in the energy of phonons, and can be explained quantitatively in the frame of the PB model.

The authors thank B. Barbara, S. Miyashita, P.C.E. Stamp and J. Brooks for fruitful discussions. This work is supported by the Alfred P. Sload Foundation and the IHRP-NHMFL-5059, and the NSF-NIRT-DMR-0506946 programs.

\end{document}